\documentclass[twocolumn,showpacs,aps,amsmath,amssymb,floatfix,preprintnumbers]{revtex4}
\usepackage{graphicx}
\usepackage{dcolumn}
\usepackage{bm}
\usepackage{longtable}
\begin{document}   

\title{Generalization of the iterative perturbation theory 
and metal-insulator transition 
in multi-orbital Hubbard bands}
\author{Takeo Fujiwara$^1$, Susumu Yamamoto$^1$, and 
 Yasushi Ishii$^2$}
\affiliation{$^1$Department of Applied Physics, University of Tokyo,   
 Tokyo 113-8656, Japan\\
 $^2$Department of Physics, Chuo University, Tokyo 112-8551, Japan}
\date{\today}
\begin{abstract}
The iterative perturbation theory of the dynamical mean field theory 
is generalized to arbitrary 
electron occupation in case of multi-orbital Hubbard bands. 
We present numerical  results of doubly degenerate E$_g$ bands in a 
simple cubic lattice. 
The spectrum shows the electron 
ionization and affinity levels of different electron occupations. 
For sufficiently large Coulomb integral, a gap opens in the spectrum 
at integer filling of electrons 
and the system becomes insulator. 
The present scheme  is  easy to combine with  
the LSDA  electronic structure theory.
\end{abstract} 
\maketitle 
The first principle electronic structure theory has been developed 
much recently in the field of strongly correlated electron systems, e.g.
the LSDA+U method,~\cite{LSDA+U} the GW-approximation~\cite{GWA} etc. .
The LSDA+U method can include the spin, charge and orbital fluctuation 
in real space, and the GW approximation is formulated to treat 
the dynamical correlation of electrons.
Even so, these are still far from the goal 
since low energy excitation of local charge fluctuation, which
plays an important role near the metal-insulator transition, 
is not properly formulated in the framework of the LSDA+U method 
or the GW-approximation.
The metal-insulator transition and the anomalous metallic phase 
is now at the stage of our understanding 
in a unified picture of both low and high energy excitations 
due to application of the dynamical mean field theory (DMFT) 
developed  by Georges and Kotliar.~\cite{DMFT-1992,DMFT-1993,RMP}

On the other hand, the DMFT might not be completely established.
Several methods have been developed in order to calculate the Green's 
function in the DMFT. 
The DMFT projects a system onto 
the single impurity Anderson model and 
this treatment is exact in the limit of the infinite dimension 
where the off-site Coulomb interaction can be neglected. 
The combination with the GW approximation is being developed 
to include the off-site Coulomb interaction.~\cite{GWA-DMFT} 
It is also tried to  extend the single impurity approximation  
to the cluster approximation.~\cite{cluster-DMFT} 
The effects of multi-orbitals were 
discussed~\cite{Rozenberg-97,Hanetal98} and  
the case of electron filling up to 1 was shown to be essentially 
the same as the case of the non-degenerate orbital.~\cite{Rozenberg-97}
The combination with the LDA 
or the LSDA is also desirable.~\cite{LDA++,Anisimovet-al,LSDA-DMFT}

To calculate the Green's functions of the DMFT, 
one can use several computational schemes such as 
the quantum Monte Carlo simulation (QMC), 
the iterative perturbation theory (IPT), 
the non-crossing approximation (NCA), 
the exact diagonalization (ED), 
the numerical renormalization group (NRG) etc..~\cite{RMP} 
QMC, ED and NRG are exact and, because of the exact 
calculation, would be limited for wide variety of applications. 
Other methods, IPT and NCA, are approximate methods and might be 
easier to combine with other calculation in 
realistic complex materials. 
The aim of this paper is to generalize the IPT to  
multi-orbital bands and to apply it to the metal-insulator
transition in the doubly degenerate E$_g$ orbitals.

We start with  the Hubbard-type Hamiltonian:
\begin{eqnarray}
 H &=&  \epsilon_d^0\sum_{jm\sigma}
       c_{jm\sigma}^\dagger c_{jm\sigma} 
       +\sum_{jm\sigma j^\prime m^\prime\sigma^\prime}
 h^{jj^\prime}_{m m^\prime \sigma}
 c_{jm\sigma}^\dagger c_{j^\prime m^\prime \sigma}        \nonumber \\
 &+& \frac{1}{2}\sum_{jmm^\prime\sigma} 
     U^{j}_{m m^\prime}n_{jm\sigma}n_{j m^\prime -\sigma} \nonumber \\
 &+& \frac{1}{2}\sum_{j\sigma m\ne m^\prime}
       ( U^{j}_{m m^\prime}-J^{j}_{m m^\prime})n_{jm\sigma}n_{j m^\prime\sigma} \ .
\label{lda1} 
\end{eqnarray}
In the following calculations, we would simplify as 
$ U^{j}_{m m^\prime}=U $ and  $ J^{j}_{m m^\prime}=J=0$, but
its generalization is just straightforward. 
The local Green's function is then defined, 
since we neglect the ${\bf k}$-dependence of the self-energy 
$\Sigma$ as usual in the DMFT, as 
\begin{eqnarray}
& & G_{m\sigma:m^\prime\sigma^\prime} (i\omega) \nonumber \\
& &= \frac{1}{V}\int d{\bf k}
 \Big[ (i\omega +\mu)\openone-\epsilon_d^0 \openone -h({\bf k})-\Sigma(i\omega)\Big]^{-1}_{m\sigma: m^\prime\sigma^\prime} \ ,
\label{L-Green}
\end{eqnarray}
where  $h({\bf k})$ is the transfer matrix in ${\bf k}$-space,
$\mu$ is the chemical potential, $\Sigma(i\omega)$ is the self-energy, and 
$\openone$ is the unit matrix.  
In case of a paramagnetic and orbitally degenerate system,  
which is the case in the present paper, 
the self-energy $\Sigma_{m\sigma:m^\prime\sigma^\prime}(i\omega)$ 
and the local Green's function $G_{m\sigma: m^\prime\sigma^\prime}(i\omega)$
are diagonal with respect to orbitals and spins and independent on them.

The local Green's function $G(i\omega)$ can be then in a form of 
\begin{eqnarray}
G(i\omega)=\bigl[ i\omega+\mu-\Delta(i\omega)-\Sigma(i\omega)\bigr]^{-1} \ ,
\label{scf-1}
\end{eqnarray} 
where $\Delta(i \omega )$ is the effective hybridization function. 
From the definition of the ``effective'' medium, one  defines 
an Green's function of the effective medium as 
\begin{eqnarray}
G^0(i\omega)=\bigl[ i\omega+{\tilde \mu}-\Delta(i\omega)\bigr]^{-1} \ .
\label{scf-2} 
\end{eqnarray} 
The second order self-energy can be defined as  
\begin{eqnarray}
 \Sigma^{(2)}(\tau) = -U^2(N_{\rm deg}-1)G^0(\tau)^2G^0(-\tau) \ , 
\label{sig0-tau}
\end{eqnarray}
where $N_{\rm deg}$ is the degeneracy with respect to spins 
and orbitals (here, in case of E$_g$ orbitals, $N_{\rm deg}=2\times 2=4$).

The IPT was developed by  Kajueter and Kotliar~\cite{Kajueter-Kotliar}
for non-degenerate orbital. 
Here, we assume a similar form of the self-energy as
\begin{eqnarray}
 \Sigma(i\omega ) = Un(N_{\rm deg}-1) +\frac{A\Sigma^{(2)}(i\omega)}{1-B(i\omega)\Sigma^{(2)}(i\omega)}
 \ , \label{IPT-1}
\end{eqnarray}
where $ n_d = \sum_{m\sigma} n_{j m\sigma} =n{N_{\rm deg}}$ and 
$n$ is actually the occupation par each orbital.
The coefficients $A$  and $B$ should  be determined under the following 
guidelines:
(1) $A$ is determined so that the self-energy $\Sigma(i\omega)$ is correct 
in the limit of $i \omega \rightarrow \infty$.
(2)  $B(i\omega)$ is determined so that the self-energy $\Sigma(i\omega)$ 
is correct in the atomic limit, {\it i.e.} $U\rightarrow \infty$.

The  exact form of the self-energy in the limit of
$i\omega \rightarrow \infty$ is generally given as
\begin{eqnarray} 
&&
 \lim_{i\omega\rightarrow \infty} 
    i\omega \{\Sigma_{m\sigma: m^\prime\sigma}(i\omega)  
 +\langle \{[H_I, c_{m\sigma}], c_{m\prime\sigma}^\dagger\} \rangle \} 
\nonumber \\
&&
 = \Big[ \langle\big\{[H_I, c_{m\sigma}],[c_{m\prime\sigma}^\dagger,H_I]\big\}\rangle 
                              -\langle\big\{[H_I, c_{m\sigma}],c_{m\prime\sigma}^\dagger\big\}\rangle^2\Big] ,  \nonumber \\   
\label{sig-1}  
\end{eqnarray}
where $H_I$ is the  electron-electron interactions. 
The brackets $[ \cdots ]$ and $\{ \cdots \}$ are the commutator and 
anti-commutator, respectively. 
In the present case, the limiting form of the self-energy is
\begin{eqnarray} 
 &&
 \lim_{i\omega\rightarrow \infty} 
 i\omega \{\Sigma_{m\sigma: m^\prime\sigma}(i\omega)
    - Un(N_{\rm deg}-1) \} \nonumber \\
 &&
\simeq   U^2(N_{\rm deg}-1) n(1-n) \, \label{sig-2} \ .
\end{eqnarray}
Here we adopted a decoupling approximation 
$\langle n_{m\sigma}n_{m^\prime\sigma^\prime} \rangle  \simeq n n$
for $(m \sigma ) \ne (m^\prime \sigma^\prime)$.
Using  the second-order perturbation, one gets a similar result as 
\begin{eqnarray}   
\lim_{i\omega\rightarrow \infty} 
i \omega \  \Sigma^{(2)}_{m\sigma: m^\prime\sigma}(i\omega)
 \simeq   U^2(N_{\rm deg}-1) n^0(1-n^0) \, \label{sig0-2}      
\end{eqnarray}
where $n^0$ is the occupation number for the effective medium Green's function 
$G^0$  and, together with ${\tilde\mu}$,
is determined so as to satisfy the Luttinger theorem.
Then the final form of $A$ should be 
\begin{eqnarray}
A = \lim_{i\omega \rightarrow \infty}
\frac{\Sigma(i\omega )- Un(N_{\rm deg}-1) }{\Sigma^{(2)}(i\omega )} 
  = \frac{n(1-n)}{n^0(1-n^0)}    \ . \label{A-limit1}
\end{eqnarray}

The required form of $B(i\omega)$ may  be 
\begin{eqnarray}
&& B(i \omega) 
  =\Sigma^{(2){\rm at}}(i \omega)^{-1}-[\Sigma^{\rm at}(i\omega)-Un(N_{\rm deg}-1)]^{-1}A   \ , 
\nonumber \\
\label{B-limit}
\end{eqnarray}
here $\Sigma^{(2){\rm at}}_{mm^\prime \sigma}$  
is the second order self-energy in the atomic limit.

The self-energy in the atomic limit can be calculated exactly, 
or even can be written down in an analytic form, through  
the atomic Green's function for the $N_{\rm deg}$-fold 
atomic level:~\cite{Hubbard-II}
\begin{eqnarray}
& & G_{m m^\prime \sigma}^{\rm at}(i\omega)  \nonumber \\
&= &  \frac{\delta_{mm^\prime}}{N_{\rm deg}}\sum_{n^\prime_d=0}^{N_{\rm deg}-1} 
\frac{(N_{\rm deg}-n^\prime_d)p_{n^\prime_d}+(n^\prime_d+1)p_{n^\prime_d+1}}{i\omega-\epsilon_d^0-Un^\prime_d+\mu}  \  \ \ 
\label{G_at-2}
\end{eqnarray}
, where $p_{n^\prime_d}$ is the  probability  
of finding an atom in a state of 
$n^\prime_d$ electrons present. 
Here the atomic levels may be given as $E=\epsilon_d^0+Un_d^\prime$ 
according to the electron occupation $n_d^\prime=0,\ 1,\ 2, \ \cdots$ 
and all configurations  are  degenerated 
when the electron occupation is identical.
If we would adopt more realistic Coulomb and exchange integral 
parameters $U_{mm^\prime}^j$ and  $J_{mm^\prime}^j$  
rather than the present simplified ones, the atomic levels 
should be split and they depend on the total angular momentum.

In order to get a rigorous expression of the self-energy, one can 
assume the limit of  the inverse temperature 
$\beta\rightarrow \infty$ with respect to $U$.  
Then the occupation probabilities of each configuration, 
for $ r<n_d<r+1$ \  ($r$ : integer number), are  
\begin{eqnarray}
p_r&=&1-(n_d-r)    \ , \nonumber \\
p_{r+1}&=&(n_d-r)  \ , \nonumber \\
p_{r^\prime} &=& 0 \ \ \ : \ {r^\prime \ne r, \ r+1} \ .
\end{eqnarray}
The resultant  expression of $B(i\omega)$ may be as follows; 
\begin{eqnarray}
&& B(i\omega)
  = \frac{i\omega-\epsilon_d^0 +{\tilde \mu}}{U^2(N_{\rm deg}-1)n^0(1-n^0)}
  \nonumber \\
&& \ \ \ + A\frac{(X-a)(X-b)}{X\{(a+b)^2-ab-U^2\}-(a+b)ab} \ , 
\label{B-form}
\end{eqnarray}
where, for $r<n_d<r+1$ ($r$ : integer number),  
\begin{eqnarray}
   X &=     & i\omega-\epsilon_d^0-Ur+ \mu \ , \ \ \ \ \ 
              q =  n_d-r   \ ,  \nonumber \\
  a+b&\equiv& \frac{U}{N_{\rm deg}}\{-q(N_{\rm deg}-1)+r\}  \ , \nonumber \\
  ab &\equiv& -\frac{U^2}{N_{\rm deg}}\{N_{\rm deg}(1-q)-r+2rq+q\} \ . \nonumber 
\end{eqnarray}
Then finally we get an  analytic expression of the self-energy of the 
present generalized IPT.
The resultant expression of the self-energy 
Eq.~(\ref{IPT-1}) with Eqs.~(\ref{A-limit1}) and (\ref{B-form}) 
are actually simple extension of the IPT ~\cite{Kajueter-Kotliar} 
and it is reduced to the known expression when 
$N_{\rm deg} \rightarrow 2$.
It is also similar to the results 
given in \cite{Anisimovet-al,LDA++} but not identical. 
\begin{figure}[t] 
 \includegraphics[width=4.4cm]{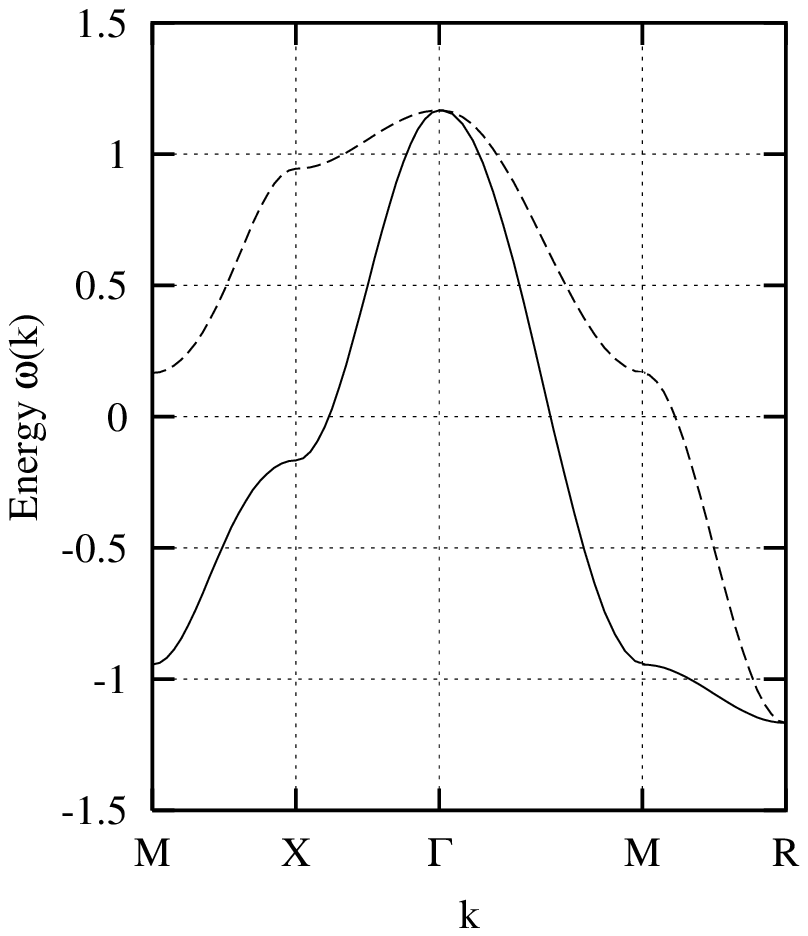} 
 \hspace{-4mm}
 \includegraphics[width=4.4cm]{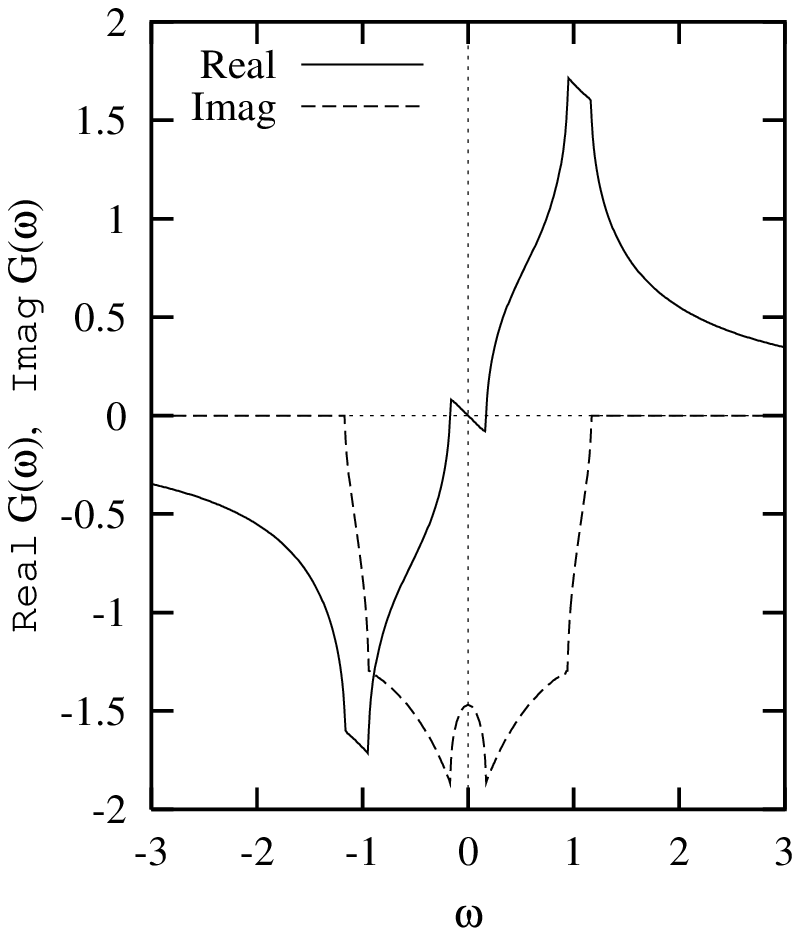} 
\caption{The energy bands $\omega({\bf k})$ 
along high symmetry lines (left) 
and the local Green's function (right) 
of Eg-bands in a simple cubic lattice.
The high symmetry ${\bf k}$-points are 
M $(\frac{\pi}{2},\frac{\pi}{2},0)$, 
X $(\frac{\pi}{2},0,0)$, 
$\Gamma$ $(0,0,0)$, 
R $(\frac{\pi}{2},\frac{\pi}{2},\frac{\pi}{2})$.
The center of the band $\epsilon_d^0=0$ and 
the transfer parameters are 
$V_{dd\sigma}={1}/{3}$, $V_{dd\pi}=-{2}/{3}V_{dd\sigma}$, and 
$V_{dd\delta}=V_{dd\sigma}/6$. 
The top $E_T$ and the bottom $E_B$ of the bands are 
$E_{T/B}=\epsilon_d^0 \pm 3(V_{dd\sigma}+V_{dd\delta})$.
The half band width is $D=3(V_{dd\sigma}+V_{dd\delta})={7}/{6}$.}  
\label{Egband} 
\end{figure}

We now present our results of the doubly degenerate E$_g$ orbitals 
($N_{\rm deg}=4$). 
We consider the Slater-Koster type tight-binding Hamiltonian 
of the doubly degenerate E$_g$ orbitals in 
a simple cubic lattice with $\epsilon_d^0=0$.  
The effective transfer integrals are assumed only between 
nearest neighbor pairs and 
$V_{dd\sigma}={1}/{3}$, $V_{dd\pi}=-{2}/{3}V_{dd\sigma}$, 
$V_{dd\delta}={1}/{6}V_{dd\sigma}$. 
The relation among $V_{dd\sigma}$, $V_{dd\pi}$ and $V_{dd\delta}$ 
is the scaling properties of bare two-center integrals 
in the LMTO method.~\cite{Andersen75} 
Here the band width is $6(V_{dd\sigma}+V_{dd\delta})={7}/{3}$ 
(the half band width $D={7}/{6}$).  
In a simple cubic lattice, there is no $dd\pi$ 
interaction within the E$_g$ orbitals and $V_{dd\pi}$ does not appear. 
Figure \ref{Egband} shows the $\omega({\bf k})$-curve 
and the real and imaginary parts of the local Green's function 
with $\Sigma=0$ and $\beta\rightarrow \infty$.

The transfer matrix $\{h({\bf k})\}$
is diagonalized at every ${\bf k}$-point.  
The local Green's functions satisfy the relation 
$G_{x^2-y^2,x^2-y^2}(i\omega)=G_{3z^2-r^2,3z^2-r^2}(i\omega)$ and  
$G_{x^2-y^2,3z^2-r^2}(i\omega)=0$ 
after the ${\bf k}$-integration within the whole Brillouin zone.    
To calculate  Green's functions, 
$8\times 343$ ${\bf k}$-points are used in the whole Brillouin zone 
and the generalized tetrahedron method for an arbitrary complex function 
is applied, which is the linear extrapolation scheme 
within a unit tetrahedron 
consisted of four ${\bf k}$-points. 
The Pad\'{e} approximation~\cite{Pade} is performed for analytic continuation 
of the Green's function 
from on the Matsubara frequency $i\omega_n$  to the  real $\omega$-axis.

\begin{center}
\begin{figure*}[t] 
\centerline{ 
\includegraphics[width=6.0cm,clip]{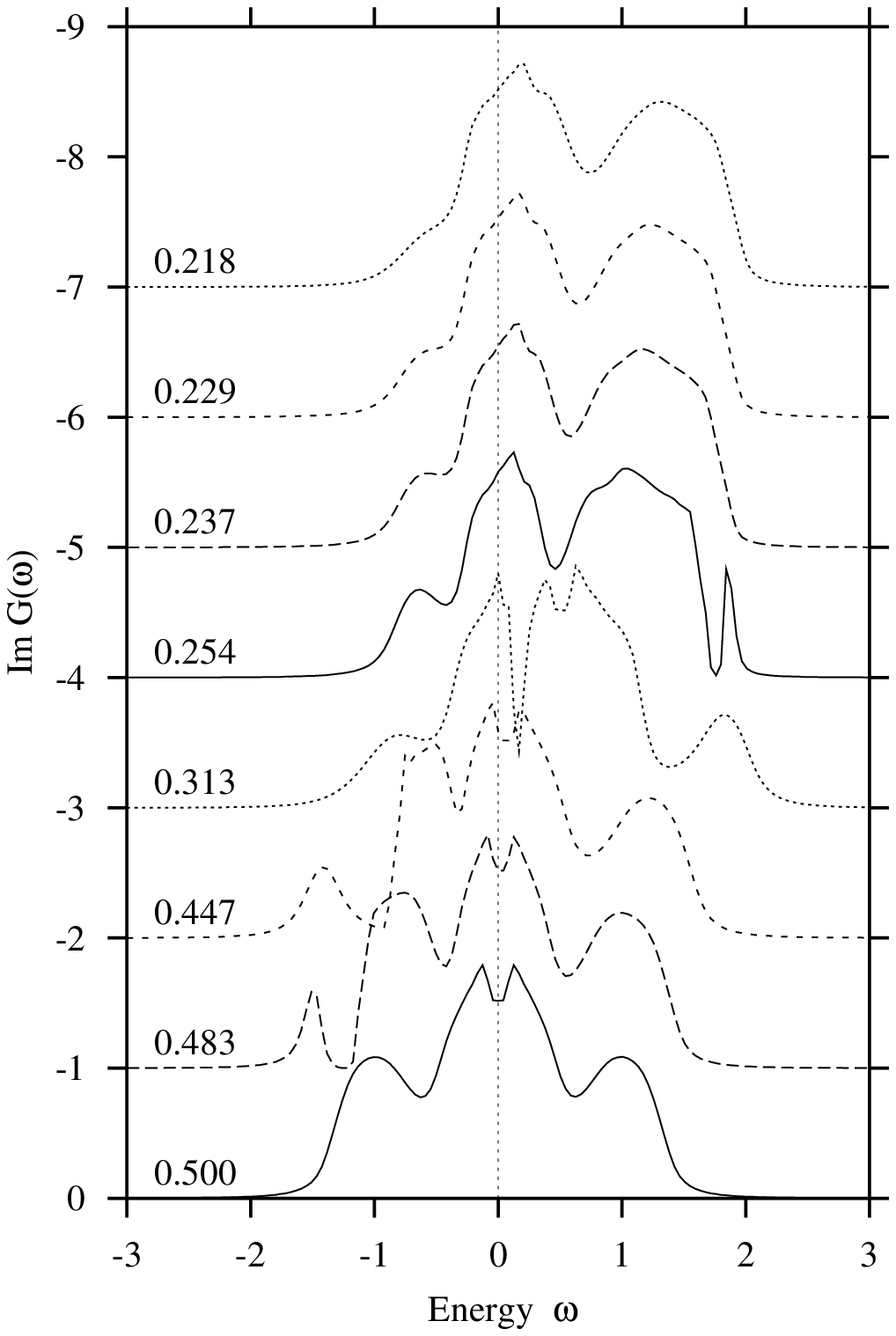} 
\hspace{-2mm}
\includegraphics[width=6.0cm,clip]{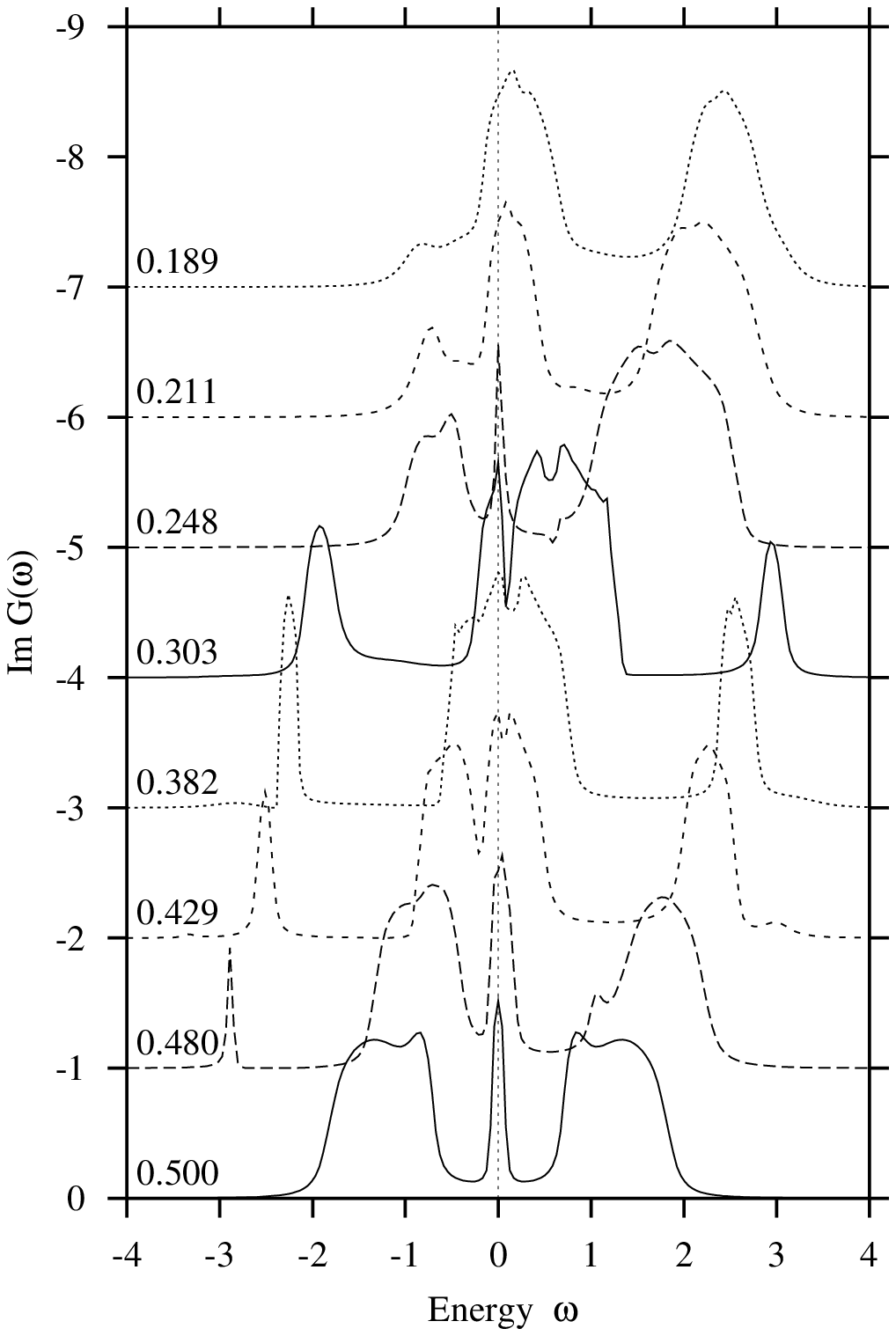} 
\hspace{-2mm}
\includegraphics[width=6.0cm,clip]{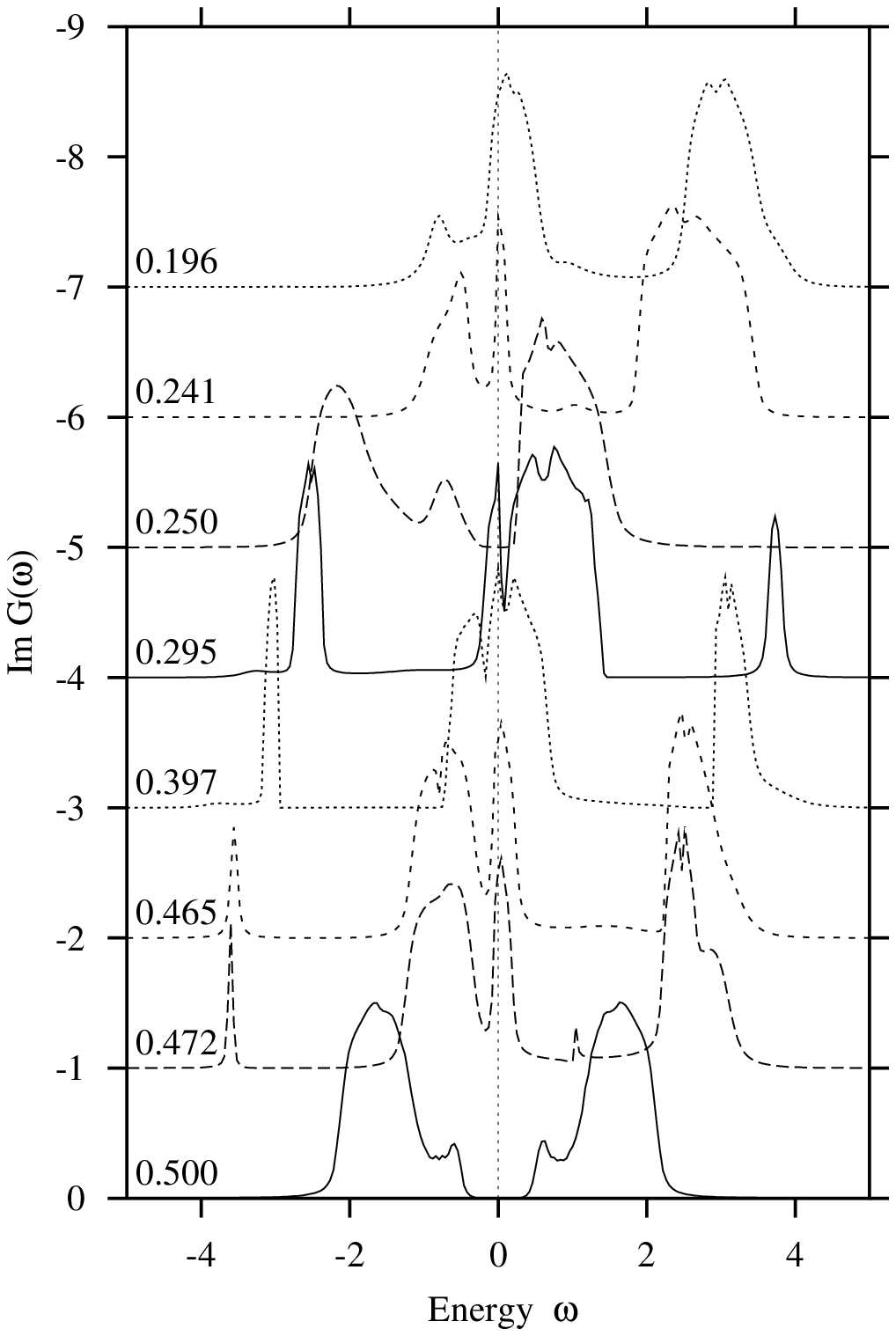}  }	           
\vspace{2mm}
{\large \bf  \hspace{1.0cm} (a) \hspace{5cm} (b) \hspace{5cm} (c) }
\caption{The electron occupation dependence of the imaginary part of the 
local Green's functions for the doubly degenerate E$_g$ band of the half 
band width $D={7}/{6}$. 
(a) $U=1.0$, 
(b) $U=2.2$ and (c) $U=3.0$. 
The numbers in the figure denotes the electron occupation $n$ for 
each spectrum. 
The inverse temperature  $\beta = 30$. 
The region $\omega<0$ is occupied and that $\omega>0$ 
is unoccupied. 
In  case of the electron occupation $n={1}/{4}$ and 
${1}/{2}$ at $U=3.0$, a gap opens  and systems are in insulating phase.}
\label{lDOS} 
\end{figure*}
\end{center}

The imaginary part of the local Green's function is  shown 
in Fig.~\ref{lDOS}, 
for $U=1.0, \ 2.2, \ 3.0$ and $\beta=30$  with electron 
occupation $0<n \le {1}/{2}$ or $0<n_d \le 2$, 
where the energy region of $\omega<0$ is occupied and that 
of $\omega>0$ is unoccupied. 
The chemical potential decreases from ${1}/{2}U(N_{\rm deg}-1)$ 
with decreasing $n$ from ${1}/{2}$.

In case of $n={1}/{2}$ or $n_d=2$ (half-filled), 
the spectrum is very similar 
to that of the case of the non-degenerate band. 
The spectrum consists of the upper and 
lower Hubbard bands with the electron-hole symmetry. 
At the critical region (here, $U=2.2$) one observes a sharp coherent peak 
at $\omega \simeq 0$. 
The change of the chemical potential with occupation is 
very rapid near $n={1}/{2}$ and $n={1}/{4}$ when $U=1.0$ and $U=2.2$ 
and it even jumps at these two occupations when $U=3.0$. 
Therefore,  the system becomes insulator 
at $n={1}/{4}$ and ${1}/{2}$ when $U=3.0$ ($U/D \simeq 2.57$).  
These occupations correspond to the cases of the total electron 
numbers  $n_d=1$ and $2$, respectively, 
and this is actually the case 
where some Hubbard sub-bands are completely filled 
and others are empty.  
The band gap of $U=3.0$  at $n={1}/{4}$ 
is narrower than that at $n={1}/{2}$ 
and then the critical $U/D$-value for metal-insulator transition 
may be smaller at $n={1}/{4}$ than that at $n={1}/{2}$. 
This prediction is consistent with the results 
by QMC and, furthermore, it is presumed that 
the critical $U/D$-value is smaller than the model 
of semicircular density of states 
in a multi-orbital case ( $(U/D)_{\rm cr}=3.1$ at $n={1}/{2}$).~\cite{Rozenberg-97}
This fact suggests that the critical $U/D$-value would depend 
upon the $\omega({\bf k})$ structure or the shape of the density of states. 
In insulating case,  $U=3.0$ with $n={1}/{4}$ and $n={1}/{2}$, 
the self-energy shows  the singular behavior 
$\Sigma \sim \frac{1}{i\omega_n}$ near $i\omega_n \sim 0$. 
On the contrary, in almost all cases, the system is metallic 
and the self-energy behaves as 
$\Sigma- Un(N_{\rm deg}-1) \sim {i\omega_n}$ which is the property of the Fermi liquid. 
The case $U=2.2$ is at the critical region of the metal-insulator 
transition, and one can see very sharp coherent peak at $\omega=0$ 
when  $n={1}/{4}$ and $n={1}/{2}$. In this region, the self-energy shows the common feature of the Fermi liquid. 

With small hole doping, the spectrum changes very drastically 
especially when $U=2.2$ and $U=3.0$. 
A  sharp peak appears at $U$ below the lower Hubbard band, whose 
width depends upon the value $U$ or the separation from the main lower 
Hubbard band.  
In this region of ${1}/{4}<n<{1}/{2}$ or $1<n_d<2$, the lower energy state is 
the mixture of the ground states of $n_d=1$ and $n_d=2$. 
Then, the ionization  and affinity levels for 
both $n_d=1$ and $2$ configurations should appear in the spectrum.  
The satellite structure of all configurations 
$n_d=0,\ 1,\ 2, \ \cdots$ could appear, 
but the intensity of satellites would depend on the energy separation. 
These satellite structure originates from  
the self-energy correctly given in the atomic limit. 

Near the occupation $n \sim {1}/{4}$, the satellite at $U$ above the upper 
Hubbard band shrinks. When the concentration $n$ crossed 
the value ${1}/{4}$, {\it i.e.} $0<n_d<1$, 
the upper satellite disappears and 
the lower satellite merges into the main (lower) Hubbard band, 
since the main contribution to the spectrum comes from 
the mixture of $n_d=0$ and 1, rather than $n_d=1$ and 2 of 
the case $1<n_d<2$.
In fact at this concentration, the chemical potential shifts 
rapidly or jumps and  the satellite grows to a main lower Hubbard 
band.  
Then the behavior for $0<n<{1}/{4}$ is rather similar to the case of 
$0<n<1$ of the non-degenerate orbital and 
the difference is only the fact that the intensity of the lower and  upper 
bands starts from the ratio of $1:3$ near $n\sim {1}/{4}$ rather than $1:1$ 
of $n \sim {1}/{2}$ in case of the non-degenerate orbit. 
 
The ${\bf k}$-dependent spectrum can be analyzed 
through the ${\bf k}$-dependent Green's function 
$G_{mm^\prime \sigma}(\omega, {\bf k})\equiv [(\omega+\mu)\openone -\epsilon_d^0 \openone -h({\bf k}) 
-\Sigma(\omega)]^{-1}_{mm^\prime \sigma} $.
The band dispersion width maybe sensitively depend upon the 
electron occupation. 
In the region of $\omega \sim 0$, the Green's function behaves as 
$G(\omega, {\bf k}) \simeq Z/\{\omega \openone -Zh({\bf k})\}$, 
where $Z$ is the renormalization factor. 
This situation may be seen, in case of the small doping from half filling.
the top of the coherent peak shows a small dip which is a 
characteristic feature of the lattice Green's function of the 
E$_g$ state.
The height of the coherent peak is almost equal to each other 
at $n\simeq {1}/{2}$ and $n \simeq {1}/{4}$. Then the renormalization factor or the effective mass may be also equal. 

In general, it may be very important to have a simple scheme for solving 
the self-consistent Green's function in multi-orbital cases, 
since further generalization is needed for  
application to realistic materials 
showing very interesting interplay of spin, charge and 
orbital ordering.
For that purpose, the present scheme is very 
practical and straightforward and, furthermore, 
the generalization for magnetically ordered systems 
and clusters containing several atoms may be quite simple. 

In conclusion, we developed a new generalization of the self-energy 
of the DMFT-IPT applicable to arbitrary electron occupation $0<n<1$ 
or $0<n_d<N_{\rm deg}$. The spectrum shows the electron 
ionization and affinity levels of different electron occupations. 
For sufficiently large Coulomb integral, 
the system becomes the insulating state at integer filling of $n_d$. 
This generalization of the IPT may be easily applicable to 
more general cases and the combination with the LSDA in  
more complex systems. 

This work was supported by a Grant-in-Aid for COE Research 
``Phase Control in Spin-Charge-Photon Coupled Systems'' and
a Grant-in-Aid from Japan Ministry of Education, Culture, Sport, Science 
and Technology. 

\end{document}